\documentclass{kapproc} 

\usepackage{graphics,graphicx}

\upperandlowercase

\begin{document}

\articletitle{A Biography of the Magnetic Field of a Neutron Star}


\author{Malvin Ruderman}
\affil{Department of Physics and Columbia Astrophysics Laboratory\\
Columbia University}
\email{mar@astro.columbia.edu}

\begin{abstract}
After some post-natal cooling, a spinning, magnetized, canonical
neutron-star (NS) has a core of superconducting protons, superfluid neutrons, and
degenerate extreme relativistic electrons, all surrounded by a thin highly conducting
solid crust. The quantum fluids are threaded by a dense array of quantized vortex-lines
which can interact strongly with a denser and much less uniform one of quantized
magnetic flux-tubes. The physics of such a core predicts the evolution of a NS's
surface magnetic field and related phenomena as the star's spin changes.  Predictions 
include changes in NS magnetic dipole moments, anomalously small polar caps in
millesecond pulsars, properties of two different families of spin-period ``glitches",
and spin-down ages much greater than true ages for some pulsars. Quantitative
model-based estimates for all of these are given.  None are in conflict with
observations.   

\end{abstract}

\begin{keywords}
Pulsar, Neutron Star, Magnetic Field, Glitches
\end{keywords}

\section{Introduction}  

There is abundant observational evidence that the magnetic
dipole-moment of a rapidly spinning neutron star (NS) evolves during a pulsar's
lifetime. Most of the relevant data  comes from observed NS-periods ($P$) and spin-down
rates ($\dot P$). With simple plausible models of NS-magnetospheres the spin-down torque on a
(presently) solitary NS depends only on the NS spin  $(\Omega   = 2\pi /P)$ and dipole
moment $(\mu   )$:
$$I\dot\Omega \sim - {\mu^2\Omega^3\over c^3}\eqno(1),$$ 
 where $I$ is the NS moment of inertia. Fig. 1 shows  NS-surface-dipole magnetic fields,
$B\equiv \mu R^{-3}$,
over a large range of pulsar spin-periods. (All pulsars are assumed to have $I=10^{45}$ 
gcm$^2$  and radii $R=10^6$	cm.)  Also shown is the typical evolution of $B$ with $P$
from the microphysics inside a canonical, rapidly spinning, strongly magnetized NS.  In
the model considered here a NS consists mainly of a sea of superfluid neutrons
together with less dense components of superconducting protons and very relativistic
degenerate electrons. (Effects from possible, much smaller, central volumes of more
exotic particles are ignored.) Surrounding this sea of quantum fluids is a thin solid
conducting crust of thickness  $\Delta \sim 10^{-1}R$ in which the NS protons clump into
conventional, but extremely neutron-rich, nuclei.  In such an object $B$ is expected to
evolve mainly because of changes in  $\Omega$. Fig. 1 shows evolutionary segments 
$a  \rightarrow b \rightarrow  c \rightarrow  d \rightarrow  e$ predicted from this
canonical NS model. The slopes of $B(P)$ in the segments $a \rightarrow  b$,  $b
\rightarrow  c$  are not significantly different from those determined directly from
certain observations of $P, \dot P$, and $\ddot P$ (Sect. 4), and from 
inferred ages of radiopulsars observed in other
$B-P$ regions. Fig. 1 segments $a^\prime \rightarrow  a$; $a^{\prime\prime}\rightarrow  b-c$; 
$c \rightarrow  d$;
$d \rightarrow  h\rightarrow   g$;  
$d \rightarrow  e \rightarrow  f$;  $d\rightarrow   h  \rightarrow q$  give more
explicit $B(P)$ predictions than what
can                                                                                                                
be inferred directly from $B-P$ data, but can be compared with other kinds of
observations (Sects. 3, 7). How this same model leads to  two different kinds of
 ``glitches" in  pulsars is discussed in Sect. 8.  In all comparisons of
observations with model-based predictions no disagreements have yet become apparent.

\begin{figure}
\centerline{\includegraphics[width=3.8truein]{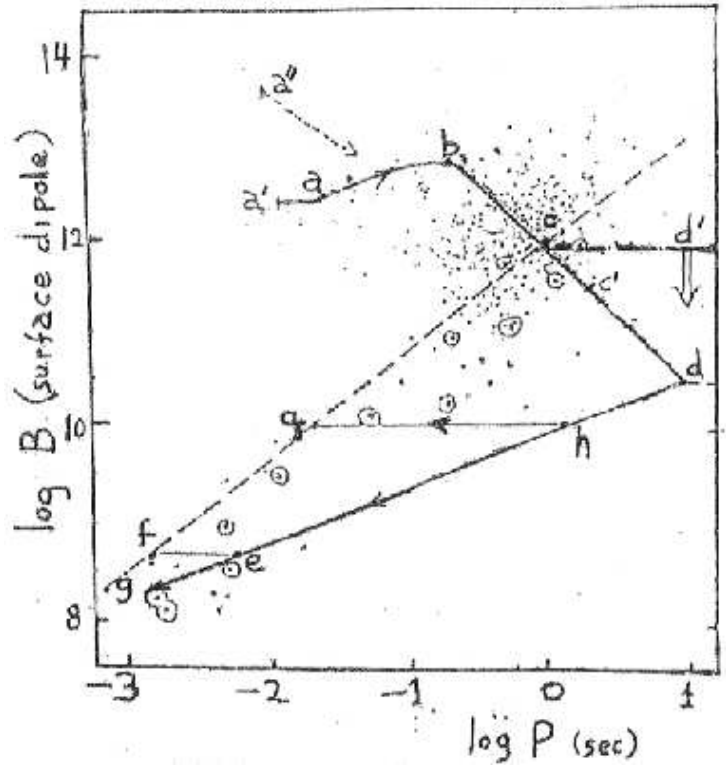}}
\caption{Dipole fields (B) of pulsars inferred from observed  $P,\dot P$, together with the $B(P)$ predicted for
a typical pulsar from the model discussed in the text. The point $c^\prime$ is about where coherent radioemission
is no longer observable. Point $d$ is for $P$ and $B$ of a solitary NS after $10^{10}$  yrs.  The millisecond
pulsar (MSP) population in the lower left corner is presumed to be populated by the evolution of some NSs off
the $bcd$ segment from accretion-induced spin-up by companions (many of which have been evaporated by the MSPs). 
After accretion stops the MSPs again spin-down like solitary radiopulsars. Open circles are pulsars in binaries 
which are not presently interacting with their companions. (The $P$, $\dot P$ of many more radiopulsars are
shown in reference [28].) The diagonal dashed line is the canonical accretion-driven spin-up line.}

\end{figure}

In almost all biographies, certainly including those of NS $B$-fields, authors and readers have
good reason to be much less confident about descriptions of earliest life (conception, infancy, and
childhood) than of adolescence, maturity, and old age.  This is even more the case here where there is
not yet a consensus about what is happening in any one of these stages. We shall begin our account of
the development of a NS magnetic field with this {\it caveat} very much in mind.                   

\section{Conception, Birth, and Infancy}

Neutron stars are believed to be born in violent
implosions of much less dense ancestors. There is no consensus about the origin of a newly formed NS's
$B (10^{12}-10^{15}  G?)$. Speculations include  

\smallskip

\noindent {(a)}  conservation during a NS's violent birth of flux already inside its ancestor (NS fluxes may
be comparable to those in magnetic white dwarfs, the toroidal field within the sun,  and, perhaps, cores of red
and blue giants);

\smallskip
\noindent {(b)} short-lived post-partem dynamos[1];

\smallskip
\noindent {(c)} field amplification in asymmetric supernova explosions;

\smallskip
\noindent {(d)} toroidal field breakout after wind-up from differential rotation imparted at birth [2];

\smallskip
\noindent {(e)} thermoelectric generation[3];

\smallskip
\noindent {(f)} exterior field reduction from burial by fall-back of some of the initially exploded matter. 
\smallskip

 Because of a NS's violent, unstable birth, the initial distribution of $B$ within a very rapidly
spinning NS is probably magneto-hydrodynamically (MHD) unstable for a time[4] 
$t_{\rm MHD} \sim 10^6 ( {10 {\rm ms}\over P} )$ sec. 
This MHD relaxation time may exceed the  ``freezing time" ($t_f \sim 10$ sec) for
neutrino emission to cool a new-born and initially very hot ($T > 10^{10}  K$) NS-crust to below the
temperature where crust-solidification begins.  Some relevant expected solid crust properties are
shown in Table 1. The main uncertainties in it are the maximum sustainable shear strain $(\theta_{\rm
max})$ and  $\Delta\theta$, the size of sudden strain-relaxation (``crust-breaking") if $Y_{\rm max}$  
is slowly exceeded.  If $t_f  >t_{\rm MHD}$ the NS begins its childhood ($t\sim 1$ yr) with relatively
small magnetic stress in its crust, but if $t_f <t_{\rm MHD},$  ${\bf j}\times {\bf B}$ forces in the
crust may sustain a surface-dipole $B$ up to ($8\pi  Y_{\rm max})^{1/2}\sim 5\times 10^{13}G$ despite
MHD relaxation to well below this for the field within most of the NS's core.  
\smallskip

\halign{#\hfil & #\hfil\cr
\noalign{{\bf Table 1. Estimates of some properties of NS-crusts}}\cr
\noalign{\hrule\vskip 1em}
     Crust thickness & $\Delta \sim 10^5$cm\cr
     Shear modulus & $\kappa \sim 10^{30}$ dyne cm$^{-2}$\cr
     Maximum sustainable shear strain & $\theta_{\rm max}\sim 10^{-3}$\cr
     Maximum averaged shear strength & $Y_{\rm max}\sim \kappa \theta_{\rm max}\Delta/R \sim
10^{26}$ dyne cm$^{-2}$\cr
     Strain relaxation for $Y > Y_{\rm max}$ & $\Delta\theta \sim
10^{-4}-10^{-3}$\cr
     Eddy current decay time &$\sim 10^7$ yrs\cr
\noalign{\vskip 1em\hrule}}
 
\section{Childhood:
$t\sim 1$ yr -- 10   yrs  (Fig. 1,  $a^\prime \rightarrow   a$; $a^{\prime\prime}
\rightarrow   b-c$)} 

About a year or so after its birth an initially very hot NS will cool its interior below the
transition temperature ($T\sim  3\times10^9  K$?) at which it becomes a proton-superconductor (p-sc).
Any magnetic field within the Type II p-sc expected within the NS core organizes itself on a
submicroscopic scale into a dense array of quantized flux-tubes (flux    $= 2\times 10^{-7}$  
Gcm$^{2}$) each of which has a radius $\Lambda \sim 10^{-11}$  cm and an interior magnetic field  $B_c
\sim 10^{15}$G. Local flux-tube area densities are huge: $n_\phi \sim 5B_{12} \times 10^{18}$cm$^{-2}$.
(If $B > B_c$, p-sc is quenched.) Submicroscopically, the $B$-field structure now becomes extremely
inhomogeneous ($B\sim 0$  between flux tubes  $5\times 10^{-1}B^{-1/2}_{12}$   cm apart, and 
$B=B_c\sim 10^{15}G$
within them).  On much larger scales the p-sc flux-tube array also varies greatly with initial
densities and twists in direction reflecting the complicated combination of poloidal and toroidal $B$
in the cooling NS core just before it began its transition to a p-sc.  Just after the transition the
previously stabilized core $B$ would be strongly out of equilibrium. The initially MHD-stabilized
configuration is based upon a compromise between minimizing  the sum of tension-energy 
$(B^2/8 \pi )$ along
${\bf B}$ and a similar $B^2/8\pi$  contribution from repulsion between field lines,  After the
transition into quantized flux-tubes the effective tensile pull per unit area jumps greatly, 
$B^2 /8\pi \rightarrow     B B_c  /8\pi$, while the repulsion between flux tubes almost vanishes
unless flux-tubes are squeezed to separation distances 
$\sim \Lambda$     so that $\bar B\sim B_c$.  Therefore, after the p-sc
transition, magnetic forces act to pull the  NS-core's magnetic field toward a new equilibrium
configuration. It is generally a much smaller one as flux-tubes try to minimize their length until
they touch each other. If this is achieved, the $\mu$   of the core-field at the surface of the p-sc
sea (its interface with the base of the NS-crust) would be greatly diminished from its value before the
transition\footnote{If,
unexpectedly, the p-sc transition is to a type I superconductor[5],  proton-supercon-\break ductivity is quenched
by a $B > B_c$   everywhere inside discrete regions very much larger than flux-tubes and vanishes
throughout the superconducting volume outside them. Then  $\overline{ B^2} > \bar B B_c$   when averaged
over large areas passing through both kinds of proton-phases. The new force distribution and its
consequences would be very similar to those following a type II transition.}.

It is difficult to calculate the time scale-needed for this new
configuration to be achieved. (Some movement of flux-tubes is by
co-advection together with their e-p-n embedding fluids. This may
involve induced flux-tube bunching and backflow between
bunches[7]. Flux-tube movement is also through those fluids). The NS's
solid, strongly conducting crust can prevent changes in magnetic field
through it for up to $10^7$ yrs as long as any shearing stress on it
at the core-crust interface $< Y_{\rm max}$. However, after the p-sc
transition, that stress may jump from $B^2/8\pi$ to $BB_c/8\pi$,
giving a reduction in the maximum possible crust-stabilized $B$ by
about 20, from a pre-childhood $B \sim 5\times 10^{13}$G to $3\times
10^{12}$G. This suggests that many pulsars may begin their childhoods
with very large dipolar Bs which may not survive this stage of their
lives. Although there is no reliable estimate of the time-scale for
this survival based upon the microphysics of all the possible
contributions to it which should be considered, an interpretation of
observed build-up of one the Crab pulsar's ``glitches" suggests about
$10^{3}$ years (Sect. 8 footnote). Then a pulsar entering childhood
with $ B\sim 5\times 10^{13}$G would spin-down to $P\sim 0.5$ sec as
its previously crust-stabilized $B$ drops to about $3\times 10^{12}$G.
Crucial evidence for such pulsars would be the observation of present
spin-down ages ($t_{sd}\equiv P/2\dot P$) of some several thousand
year old pulsars which are about 20 or so times greater than their
true ages, those inferred from the ages $(t_{snr})$ of the supernova
remnants in which these NS's are still embedded.  This large reduction
would, of course, become smaller if B before the p-sc transition were
smaller, and could disappear if that $B < 3\times 10^{12}$ G
($a^\prime$ in Fig. 1).  Fig. 2 compares spin-down and
supernova-remnant ages of Galactic pulsars for which both have been
reported. The two pulsars with the largest differences, 1E1207 and
J1952, have present magnetic dipole fields of about $3\times 10^{12}$G
and present spin-down times about $10^2$ times longer than
their true ages.  (One alternative, and presently more common,
explanation of this discrepancy is that many pulsars are born with
such very long spin-periods that they have not spun-down much from
those periods in their first $10^4$ yrs.)

\begin{figure}
\centerline{\includegraphics[height=2.6truein]{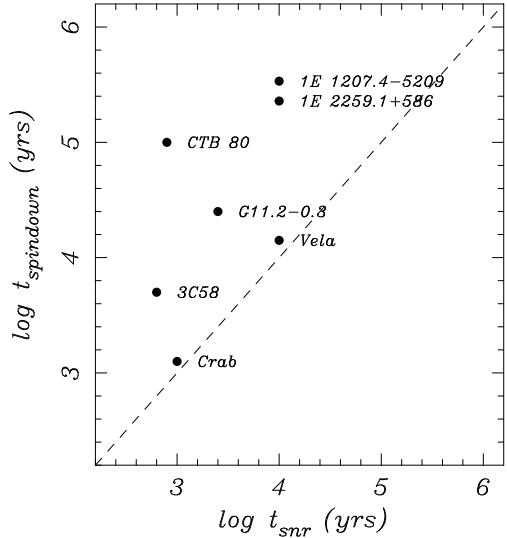}}
\caption{Observed spin-down ages $(t_{spindown} \equiv P/2\dot P)$ of some young NSs together with estimates of their true ages from
those of the supernova remnants in which they are embedded, or from historical     records of  associated
supernova explosions.}
\end{figure}

\section{Adolescence: $t\sim 10^3  -  10^4$yrs; Crab-like pulsars (Fig. 1 $a \rightarrow  b$)}

About $10^3$ yrs after their violent births, NSs will have cooled below the transition temperatures
(several $\times 10^8$K) to neutron superfluidity $(n-sf)$. An initially nearly uniformly rotating NS
neutron-sea then changes its fluid rotation pattern by establishing  a nearly parallel array of
quantized vortex lines aligned along the spin-direction. These vortex-lines, with an area number
density                                                                                                                       
$n_v \sim 10^5  (\Omega  /\Omega_{Crab})$cm$^{-2}$,  pass through the the hugely more abundant, 
curved, twisting, magnetic flux-tubes already formed in the p-sc during early NS childhood.  When the
neutrons in a NS core spin-down(-up) these corotating n-vortex-lines move away from(toward) the NS
spin-axis.  Vortex lines, parallel to the spin-axis but displaced from it by  $r_\perp$, move out
with an average velocity  
$v = -\dot\Omega   r_\perp/2\Omega \sim 10^{-5} \times$ (10$^3$yrs/NS age) cm s$^{-1}$. Because of the
strong velocity-dependence of the neutron-proton interaction, segments of n-vortex-lines and
p-flux-tubes also interact strongly with each other when they are within a distance $\Lambda$.
Therefore the moving n-vortex-lines of a NS whose $\Omega$   is changing must either cut through or
carry along with themselves the geometrically complicated flux-tube array which they thread.  Which
occurs, depends on the magnitude of the Resistance of various regions of the flux-tube array to being
moved through the embedding seas[6], MHD instabilities which can result in flux-tube bunching[7],
possibilities for advective movements of flux-tubes together with their embedding electrons and
protons, the force just before cut-through of the vortex-flux-tube interaction, and the possibility of
temporary anchoring of the core's flux-tubes by the solid crust. Both frictional drag and inelastic
cut-through generate heat within the core, Therefore thermal x-ray emission bounds for spinning-down
NSs  (especially where they exceed neutrino-emission in NS cooling) give an empirically determined
limit to both. Consideration  of the above supports the following comments.

\smallskip
\noindent
{a)} Characteristic n-vortex velocities and the induced flux-tube array velocities are small
enough in the cores of adolescent pulsars  (spin-down age $>10^3$yrs) that cut-through of
flux-tubes by moving vortex-lines is expected to be unimportant in them. 

\smallskip
\noindent
{b)} This conclusion is more firmly based for older pulsars and quite compelling for those whose
spin-up (-down) ages exceed $10^5$yrs.  (Even if the p-sc is type I  in some of the core region,
there will be such strong forces there between n-vortex lines and the field boundary regions of
thickness $\Lambda$ in which $B$ goes from $\sim 0$ to $>B_c$,  that slowly moving vortex-lines resist
passage through such boundaries and move the $B$-field regions with themselves.)                                                                                                        

\smallskip
\noindent
{c)} The crust is not strong enough to prevent the surface $B$ of a NS from following vortex-array
controlled flux-tube movement near the base of the crust as long as $B > 10^{12}$G and  $P < 1$
sec. Outside this $B-P$ range the conducting crust could delay this for about $10^7$yrs,
but not prevent it (cf. the discussion on very short delays in Sect. 8).  
 
\smallskip

\noindent    
We consider next observational consequences of a very simple evolutionary model expected  to have
validity for magnetic field evolution in adolescent NSs and, with even more confidence, for such
evolution in more slowly spinning-down (-up) older NSs: {\it the magnetic dipole field on the surface
of such NSs follows that of the core's p-sc flux-tube array near the base of the crust. The movement of
that array is, in turn, controlled by the expanding (contracting) n--sf vortex-array of the
spinning-down(-up) core n--sf.}
From here on much of our discussion follows what is in the published
literature, supplemented by new supporting x-ray data from the XMM and Chandra satellites. It will
therefore be rather abbreviated with more detail and references available elsewhere[6,7].
In the adolescent (Crab-like) pulsars $r_\perp \propto    P^{1/2}$  until 
$r_\perp$  reaches the NS radius $R$.
The predicted evolution of $\mu_\perp$ (the component of ${\bf \mu}$
   perpendicular to $\bf \Omega$) is then
particularly simple.  Models which attribute spin-down mainly to the Maxwell torque,  
$\mu^2_\perp\Omega^3/c^3$, have
a ``spin-down index" $n\equiv -\Omega\ddot \Omega \dot\Omega^{-2} = 3 -
{2\dot\mu_\perp\Omega\over \mu_\perp\dot\Omega} = 2$
 as  long as all $ r_\perp\propto   P^{1/2}$.  As  $r_\perp$  reaches $R$ for a significant fraction of
the vortex lines, $n$ grows from 2 toward 3. It reaches 3 when flux-tubes cannot move out further. 
This behavior is shown in the upward moving $a\rightarrow  b$ segment of Fig 1. and $n$ values
observed in some adolescent pulsars (for which $\ddot P$ has been measured):  $n=2.5$, Crab[8];  $n=2.8$, PSR
1509[9];  $n=1.8$, PSR 0540[10];  $n=2.9$, PSR J1119[11].
 (For a different model for $2<n<3$ in which  the effective I varies cf. ref. 30).)
                              
\section{Maturity:  ages $10^4  - 10^6$yrs; Vela-like and older pulsars (Fig.1  $b \rightarrow  c$)}

As
flux-tubes are pushed out of the  p-sc  core  by the core's expanding n-sf vortex-array,  North and
South poles at the surface will ultimately reconnect. (cf the discussion in Sect.8 on ``giant
glitches".) Thereafter,  $\mu_\perp$   typically decreases as $P^{-1}$, which gives an average
spin-down index $\langle n\rangle =5$.  (This is because the part of the core's flux, which has not yet
been pushed out to  $r\sim R$  where it reconnects, is the source of the NS's remaining surface dipole
$B$. That remaining flux is proportional to the  remaining number of core n-sf vortex-lines.)  
Differences between spin-down ages and kinematically determined ones of
observed pulsars
around the $b\rightarrow   c$ segment of Figure~1 implies $ n=4.5\pm 0.8$[12]   The $n$ at a particular $P$ for any one pulsar cannot
be predicted {\it a priori} without detailed knowledge of the NS core's magnetic field structure\footnote{For
example, N and S surface polar caps might be connected by pushed core-flux just below the surface.
This pulls them toward each other until they ultimately reconnect $(n>3)$, or first pulls them apart
$(n <3)$ until they are on opposite sides of the star . In the latter case further pulling  will then
bring them closer $(n>3)$ until reconnection is finally achieved. Alternatively, an initially 
complicated surface field could have many N and S pole regions. Then the vector sum which gives the net
dipole moment may be either decreased $(n<3)$ or increased $(n>3)$  by reconnection of any one pair. 
Only after long evolution would $\langle n\rangle \sim 5$ be realized (cf the reported $n\sim 1.4$ for
Vela[13].)}.

\section{Old age (Fig.1, $c \rightarrow c^\prime \rightarrow d; c 
\rightarrow d^\prime \rightarrow d$)  }        

The point $c$ in Fig. 1 is near where the maximum
expected magnetic shear stress on the crust's base no longer exceeds the crust's yield strength  
$(BB_c/8\pi   < Y_{\rm max})$. The evolution of surface $B$ beyond $c$ depends on time scales. The core's surface
$B$ should follow the trajectory $c \rightarrow  c^\prime \rightarrow  d$, but the crust's surface $B$ would now
follow it only after ten or so million years, when crustal Eddy currents have died out (or perhaps earlier if
plastic flow in the crust has allowed sufficient reduction of its ${\bf j}\times {\bf B}$ stresses). This
time-lag is plausible as the reason why some x-ray pulsars , NSs spun-down relatively rapidly by binary
companions (e.g. to
$P\sim 10^3$ sec for Vela X-1), can temporarily maintain surface dipole $B\sim 10^{11}$G.
 
The deathline of a solitary NS as an observable radiopulsar is expected at $P\sim$ several seconds and
$B\sim 10^{11}$G, but spin-down would continue.

\section{Resurrection of some $10^8  -10^{10}$ yr old  NSs (Fig. 1 $d \rightarrow  h \rightarrow  g$; $d
\rightarrow  h \rightarrow  e \rightarrow  f$; $d \rightarrow  h \rightarrow  q$; $d \rightarrow  c$)}

Some dead pulsars in binaries will be spun-up by accretion from companions which have evolved,
orbitally or in size, to fill their Roche lobes. (Before or during this phase, the interaction with
the companion may first have given the NS larger additional spin-down with accompanying reduction in B
than would have been the case if it was solitary.) Such a genesis by accretion from a low mass
companion (LMXB) is widely proposed as the origin of the millisecond radiopulsars (MSPs) in the lower
left corner of Fig.1. The superfluid-vortices's radial velocities within a NS being spun-up to a
millisecond period in an LMXB ($\sim 10^{-9}$cm sec$^{-1}$) are so small that flux-tube movement which follows
it seems inescapable. The very slow inward movement of the n-sf votices squeezes all magnetic flux inward
with it toward the NS spin-axis. Evolutionary tracks for surface-dipole $B$ in Fig. 1 depend upon the
initial B-field configuration at the beginning of the long, slow spin-up[6,14,15,16]: 
  
($d \rightarrow   c$) -  {\it N and S polar caps are in opposite spin-hemispheres}. The final dipole moment is
almost aligned and somewhat bigger than it was initially. 
  
($d \rightarrow   g$) - {\it  N and S polar caps are in the same spin-hemisphere}. They are then squeezed
together near the spin-axis to form an orthogonal rotator  $(\mu \perp \Omega)$  whose  
$\mu$    is reduced by
a factor $(P_g  /P_d  )^{1/2}\sim  10^{-2}$. Spin-up alone, by bringing the N and S magnetic poles so close
together, reduces $\mu$   to the small value essential for accretion to spin-up the NS to a MSP (by approaching
the limiting accretion spin-up line of Fig. 1).
   
($d \rightarrow   h  \rightarrow q ; d \rightarrow  e  \rightarrow f)$ -  {\it most, but not all, flux from
either spin-hemisphere returns to the  NS surface in the same hemisphere from which it came.}
 Continued spin-up
reduces the orthogonal component,  $\mu_\perp \propto  (P)^{1/2}$,  together with a slightly increased,
aligned $\mu_\parallel$. When these two components of the total dipole approach equal magnitudes at, say, $h$
(or $e$), the total dipole moment (and surface dipole $B$) can no longer be strongly reduced by further spin-up
since it diminishes only the orthogonal component. The initial conditions needed for accretion-induced spin-up
into the extreme lower left corner of Fig. 1 from such $d \rightarrow  h \rightarrow  e$  strongly suggest that
the exceptionally fast MSPs they evolve into should usually have a surface magnetic field resembling that from an
orthogonal dipole positioned on the NS spin-axis at the interface between the NS's perfectly diamagnetic p-sc
core and the bottom of its crust (Fig. 3b).  A second large MSP family there  should consist of almost aligned
rotators (less frequently observed since their radio beams are directed so close to the NS spin-axes) whose    
$\mu_\perp$have been ``spin-up squeezed" to negligible strength. Their surface-field configuration should then
closely resemble that from  a N(S)  pole on the spin-axis where magnetic field leaves the diamagnetic core in
the upper spin-hemisphere and an equal strength S(N) pole where it re-enters the core in the   lower
spin-hemisphere (Fig. 3a). In both MSP families the size of the polar caps at the crust core interface is
expected to be small.  [Polar cap radii there $\sim (P/P_d)^{1/2}   R \sim  10^4$cm.]
 There is strong observational support for both of these two MSP families.

\begin{figure}
\centerline{\includegraphics[width=3.5truein]{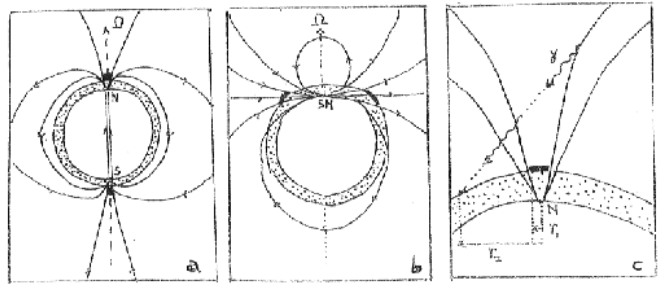}}
\caption{Magnetic field configurations of strongly spun-up millisecond pulsars (MSPs):  (b) field configuration
after prolonged spin-up when magnetic poles were in the same spin-hemisphere; (a) final configuration when N and
S poles were in opposite spin-hemispheres or had a more complicated distribution with many poles but the spin-up
was extremely great; (c) a magnified view of (a) around the North polar cap. Indicated in (c) is a gamma-ray of
the curvature radiation from an extreme relativistic lepton moving in from an accelerator along an open field
line. Where it hits the surface there is a hot polar cap (very black in the figures). Energetic curvature
radiation disappears near the polar cap because of the disappearing curvature of open field-lines near the
essentially isolated pole.} 
\end{figure}

\smallskip
\noindent
{(a)}  An exceptionally large fraction of the most rapidly spinning MSPs in the Galactic disk fit orthogonal
rotator criteria of two sub-pulses of comparable strength separated by around 
$180^\circ$   in phase[15,17]. Of the 6 MSPs 
reported to have x-ray pulses as well as radio-emission[18], 5 are consistent with orthogonal rotators. This
same orthogonality criterion is met in only about $10^{-2}$ of the rest of the radiopulsar population. 

\smallskip
\noindent
{(b)}  The very
special predicted $B$-field structure in the near-magnetosphere from ``spin-up flux-squeezing" into an orthogonal
rotator gives just the radio-subpulse polarization properties, and their frequency dependence[15], observed in
the first,  fastest MSP, PSR 1937 $(P=1.6\ ms)$. 

\smallskip
\noindent
{(c)} The sixth x-ray identified radio-MSP, PSR 0437, has a huge
radiopulse-width ($\sim 270^\circ$). This and the structure of its radiopulse-polarization[19] strongly support
its categorization as a nearly aligned rotator. 

\smallskip
\noindent
{(d)}  Just below the accretion spin-up line of Fig.~1 is where a
large fraction of nearly aligned pulsar candidates are observed.   

\smallskip
\noindent
{(e)}  The aligned MSP, PSR 0437 ($P\sim 6$ ms),  would
have a surface polar cap area  $A_{p c} \sim \pi\Omega    R^3 /c \sim 10^{11}$cm$^2$   for a conventional
central dipole field or that from a uniformly magnetized core. This $A_{pc}$   should also be the emission area
of blackbody radiation x-rays sustained by backflow of extreme relativistic particles down onto the polar cap
from the pulsar's open field line particle accelerator(s).  The expected blackbody radiation is indeed observed,
but its emission area is only $4\times 10^8$cm$^2$, and it is surrounded by a comparably luminous but cooler 
blackbody annulus $2\times 10^5$cm away[20].  These two observed features are hard to understand with
conventional models of a NS's surface magnetic field. They are, however,  just what are predicted on the surface
of a strongly spun-up, flux-squeezed, aligned MSP[14].  In that model open field line bundles span very much
smaller polar cap areas on the crust surface  (just above  each of the two core poles)  than  polar cap areas in
models  with conventional B-field configurations. The new predicted polar cap area
$\sim \pi\Omega  R\Delta^2  /c \sim 10^9$cm$^2 \sim 10^{-2}  \times$ that from
central dipole models.  For PSR~0437 the predicted polar cap radius , 0.17~km, is a tenth the canonically
estimated one and consistent with the 0.12~km deduced from observations. Curvature radiation from extreme
relativistic particle inflow onto such polar caps is strong along almost all of an incoming particle's
trajectory and would heat up a large surface area extending far from the polar caps. However, with the special
squeezed flux geometry of Figs.~3a,c such strong curvature radiation sources should disappear 
above the polar cap where the local $B$-field lines lose their curvature. Strong curvature radiation heating of the NS surface
outside the polar cap itself should then only be important beyond about 2~km away, in agreement with the inner
radius of the reported hot annulus.

\bigskip
Up to this point all of the agreement between the simple spinning-NS model predictions and related observations
have involved relatively slow, time averaged, changes in B at the surface of a NS crust as it responds to
changing spin of a NS core's quantum fluids. We turn now to a consideration of other observations which test
expectations of the detailed way in which such crustal field changes are accomplished.

\section{Small sudden changes in $B$ through overstressed crusts:  pulsar ``glitches"}

Moving core flux-tubes continually build up stress in surrounding conducting crust which anchors the
B-field that traverses it. If this stress grows to exceed the crust's yield strength $(Y_{\rm max})$,
subsequent relaxation may, at least partly, be through relatively sudden crustal readjustments     
(``crust breaking"). Such events would cause very small spin-ups(-downs) in spinning-down(-up)  NSs
(spin-period ``glitches"). The above model for the evolution of a core's flux-tube array in adolescent
and mature pulsars suggests glitch details in such pulsars similar to those of the two observed glitch
families: Crab-like glitches (C) and the very much larger giant Vela-like ones (V).   

\smallskip
\noindent
{\sl a)  Crab-like
(C) glitches.} In both adolescent and mature  pulsars an expanding quasi-uniform n-sf vortex-array
carries a p-sc flux-tube array outward with it. If growing flux-tube-induced stress on the crust is
partly relaxed by ``sudden" outward crust movements (of magnitude $s$) where the stress is strongest (with
density preserving backflow elsewhere in the stratified crust) the following consequences are expected:

\smallskip\noindent
{(1)} a ``sudden" permanent increase in $\mu_\perp$, spin-down torque, and   $|\dot\Omega | :         
{\Delta\dot\Omega\over \dot\Omega} \sim s/R \sim \Delta\theta$
 (strain relaxation) $< \theta_{\rm max} \sim  10^{-3}$. (This is the largest non-transient
fractional change in any of the   pulsar observables expected from  breaking  the crust.) A permanent
glitch-associated jump in NS spin-down rate  of this sign and magnitude ($\sim 3\times 10^{-4}$) is
indeed observed in the larger Crab glitches[21,22].   

\smallskip\noindent
{(2)} a ``sudden" reduction in shear stress on the crust by the flux-tubes below it. Its estimated 
magnitude is $(BB_c  /8\pi  )(s/R)$. This is also the reduction in pull-back on the core's expanding n-sf vortex
array by the core's p-sc flux-tube array which it tries to drag with it. 
The n-vortices therefore move out to a new equilibrium position 
where the Magnus force on them  is reduced
by just this amount. The  high density $(\rho)$ n-sf, therefore, spins down a bit. All
the (less dense) charged components of the NS (crust, core-p and-e) together with the n-vortex array 
must spin-up much more. (The total angular momentum of the NS does not change significantly in the brief time
for development of the glitch.) A new equilibrium is established in which the charged components (all that is
observed, of course, is P of the crust's surface)  have acquired a
$$ {\Delta\Omega\over \Omega} \sim {BB_c\over 8\pi \rho R^2\Omega^2} (s/R) \sim
10^{-4} \left( {\Delta\dot\Omega\over \dot\Omega} \right).\eqno(2)$$                           
Crab-glitch  ${\Delta\Omega\over \Omega}$       with magnitudes $\sim 3\times 10^{-8}$    and  
${\Delta \dot\Omega \over \dot\Omega}   \sim 3\times 10   ^{-4}$ 
are observed. So are many much smaller glitches, and the proportionality factor of Eqn. 2  holds approximately
for them as well[22]\footnote{One of the larger Crab glitches has been observed throughout its early
development[23].  It took a time 
$8\times 10^4$  sec
for the observed  $\Delta\Omega$     to rise to its full glitch value. One possible interpretation of this delay
is that this is the time it takes for the suddenly unbalanced force on the core's n-vortex array to drag this
array to its new equilibrium position. The main drag retarding such a repositioning would be from moving the
flux-tube array which is forced to co-move with the n-sf  vortices. If the very small speed with which the
combined vortex-line - flux-tube arrays responds ($\sim 10^{-8}$cm/sec) is assumed proportional to the small
unbalanced force exerted on them just after the crust-breaking, this would scale to about $10^3$   yrs for
flux-tubes to move to their new very distant equilibrium positions when subject to the much stronger forces on
them discussed in Sect. 3.}.

\smallskip      
\noindent {\sl b) Giant Vela-like (V)glitches.}  
A second V-family of glitches differs from that of Crab-like
ones (C) in several ways[29].  (1) $(\Delta\Omega/\Omega)_V \sim 10^2\times (\Delta\Omega/\Omega)_C$. 
(2) V-glitches
develop their $\Delta\Omega$     in less than $10^2$sec.: the $\Delta\Omega$ of a V-glitch is already decreasing in
magnitude when first resolved[24], while C-glitches are still rising toward their full 
$\Delta\Omega$   for almost $10^5$  sec[23].  (3)
V-glitches are observed in mature pulsars (mainly, but not always  in Fig. 1 along $b\rightarrow    c$)  while C
-glitches are observed in both adolescent and mature pulsars.  (4)  Eq. 2 for C-glitches would greatly
overestimate 
$(\Delta\dot\Omega    /\dot\Omega)$ for
V-glitches.   The existence of a second glitch family, with V-properties, should  result from a  different kind
of vortex-driven flux-tube movement in a NS core. If there were no very dense, comoving, flux-tube environment
around them, outward moving core-vortices could smoothly shorten and then disappear entirely as they reached the
core's surface at its spin-equator. However, the strongly conducting crust there resists entry of the flux-tube
array which the vortices also bring  with them to the crust's base. This causes a pile-up of pushed flux-tubes
in a small equatorial annulus, which opposes the final vortex-line disappearance. The final vortex-line
movement in which they vanish occurs either in vortex-line flux-tube cut-through events, or, more likely, in a
sudden breaking of the crust which has been overstressed by the increasing shear-stress on it from the growing
annulus.  Giant V-glitches are proposed as such events, allowing a sudden reduction of part of this otherwise
growing annulus of excess angular momentum  and also some of the magnetic flux trapped within it. These would
not begin until enough vortex-lines, initially distributed almost uniformly throughout the core, have piled up in the
annulus to supply the needed stress. This happens when adolescence is completed, i.e.~when  crust B-field
reconnection (maturity) begins. The P at which this occurs may vary very considerably among pulsars, depending on
their childhood B and its history. If crust-breaking displacements in such events involve crust movements of
about the same size as those in the largest C-glitches $(\vert s \vert \sim 3 \times 10^2{\rm   cm})$, and these
crust movements (with their associated expulsion of flux and its ultimate reconnection) are responsible for the
average decrease in magnetic moment of mature pulsars (Fig. 1 $b \rightarrow  c)$, then the interval between
V-glitches 
$$ \tau_g \sim 3\times 10^{-4} \left( {\Omega\over \dot \Omega}\right) \vert n-3\vert^{-1}.\eqno(3)$$
For the Vela pulsar this gives an average interval between giant glitches of about 4 years, compatible to that
observed. [Depending upon the sign of s, Vela might have 
$n\sim 3+2 =5$ or $3-2 =1$ (cf. the footnote in Sect 5.).
The $\tau_g$   from Eq.3 is the same.]  If the ``Big Glitcher"[25], PSR J0537, with its observed 
$n\sim 7$ and $t_{sd}   =5\times 10^3$  
yrs, is already a ``mature" pulsar, its predicted  $\tau_g   =0.7$ yrs. This is in reasonable agreement with its
observed  $\tau_g =0.4$ yrs.  
   
The density of excess angular momentum and flux-tubes stored in the annulus is limited by the strong repulsion
among flux-tubes when they become so closely packed that average $B$ among them approaches $B_c$.   
The annulus builds up to a volume $V_A$   where the growing stress from its coupling to the NS crust reaches the
crust's yield-strength.  This
new crust-breaking epoch is reached after very substantial spin-down, from early adolescence where vortex-lines
first develop and start their outward movement to the onset of ``maturity" where giant glitches and 
$\langle n\rangle \sim 5$ begin.
Subsequent spin-down continues forced entry into this annulus of small stretches of vortex-lines together with
those parts of accompanying flux-tubes they bring into the annulus with them.  But new inflow is now  balanced
by glitch-events which give a comparable loss of flux and angular momentum from the annulus. (The annulus does
not grow larger after the crust's yield-strength is reached.)  For a constant density neutron star core and 
$ t_{sd}  
/\tau_g$   glitches in a spin-down time, each glitch would give a relative jump in observed pulsar
spin                                                                                                                             
$$ {\Delta\Omega\over \Omega} \sim \left( {5V_A\over 2V_{NS}} \right)
 \times \left({\tau_g\over t_{sd}}\right).\eqno(4)$$
with  $V_A/V_{NS}$   the ratio of the small annulus volume $(V_A)$ to the volume of the whole NS core 
$(V_{NS}   )$.
Unfortunately the present value of $V_A  /V_{NS}$  depends on unobserved features in the development of a NS's
core B-field during infancy and childhood. If, for example, that field at the beginning of adolescence is near
the surface dipole field $B(a) \sim 2\times 10^{12}$G, and the total volume of the core's flux-tubes is
conserved as they are pushed outward and compacted into $V_A$, then $V_A  /V_{NS} \sim B(a)/B_c 
\sim 2\times 10^{-3}$.
Probably more realistically, as discussed in Sect.3,  $B$-evolution in childhood suggests                        
a final core $B$ typically about an order of magnitude larger than the surface dipole when childhood ends. But,
during subsequent outward movement and compactification  of the core's flux- tubes, it is likely to be their total
number rather than their total volume that is conserved. If so,  
$V_A  /V_{NS}  \sim [10 B(a)/B_c  ]  \sim 3\times 10^{-3}$. It is
fortuitous that both estimates  for the  $V_A  /V_{NS}$   ratio in  Eq. 3  happen to agree.  With it and Vela
pulsar parameters  $\tau_g  = 3$yrs,  $t_{sd}   = 10^4$yrs,  the predicted  
$(\Delta\Omega/\Omega) \sim 2\times 10^{-6}$, typical of that
of a giant glitch in Vela. For the ``Big Glitcher" the predicted $(\Delta\Omega   /\Omega  ) =7\times
10^{-7}$, comparable to the $4\times 10^{-7}$    of its observed ones[25].    
  
 This completes our brief biography of a canonical (usually solitary) pulsar's magnetic field through all phases
of its life. Different kinds of observations, many of which would otherwise seem puzzling, all give considerable
support for a very simple model in which the biography of a NS's magnetic field is closely and simply tied to
the history of it's spin\footnote{This glitch
model is quite different from the presently widely applied one in which some of the $n-sf$ vortices which are
located inside the NS crust and are normally pinned to crust nuclei, collectively un-pin and move outward[26,31].
This sudden movement of crust-vortices reduces the crust's n-sf angular velocity and spins-up the rest of
the NS (cf. ref. 27  for a criticism of this as a basis for a model for giant glitches). In the model proposed
here these crustal
$n-sf$ vortices do not play an important role in triggering a glitch, but may have observable consequences in
post-glitch healing.}

\begin{acknowledgments}

 I am happy to thank E.~V.~Gotthelf, J.~P.~Halpern, P. Jones, J. Sauls, J. Trumper, and
colleagues at the Institute of Astronomy (Cambridge) for helpful discussions.
\end{acknowledgments}

\end{document}